\documentclass[fleqn,12pt,twoside]{article}
\usepackage[headings]{espcrc1}

\readRCS
$Id: espcrc1.tex,v 1.2 2004/02/24 11:22:11 spepping Exp $
\usepackage{graphicx}
\usepackage[figuresright]{rotating}
\begin{document}
\newcommand{\ttbs}{\char'134}
\newcommand{\AmS}{{\protect\the\textfont2
  A\kern-.1667em\lower.5ex\hbox{M}\kern-.125emS}}

\hyphenation{author another created financial paper re-commend-ed Post-Script}

\title{Heavy Ion Collisions at Relativistic Energies: Testing a Nuclear Matter
at High Baryon and Isospin Density}
\author{M.Di Toro\address[lns]{Laboratori Nazionali del Sud INFN, I-95123 
Catania, Italy,\\
and Physics-Astronomy Dept., University of Catania}%
\thanks{ditoro@lns.infn.it},
M.Colonna\addressmark[lns],
G.Ferini\addressmark[lns],
T.Gaitanos\address[muenchen]{Dept. f\"ur Physik, Universit\"at M\"unchen, 
D-85748 Garching, Germany},
V.Greco\addressmark[lns],
H.H.Wolter\addressmark[muenchen].}

\runtitle{Testing a Nuclear Matter...}
\runauthor{M.Di Toro}

\maketitle

\begin{abstract}
We show that the phenomenology of isospin effects on heavy ion reactions
at intermediate energies (few $AGeV$ range) is extremely rich and can allow
a ``direct'' study of the covariant structure of the isovector interaction
in the hadron medium. We work within a relativistic transport frame,
beyond a cascade picture, 
 consistently derived from effective Lagrangians, where isospin effects
are accounted for in the mean field and collision terms.
Rather sensitive observables are proposed from collective flows
(``differential'' flows) and from pion/kaon production ($\pi^-/\pi^+$, 
$K^0/K^+$ yields). For the latter point relevant non-equilibrium effects
are stressed.
The possibility of the transition to a mixed hadron-quark phase, 
at high baryon and isospin density, is finally suggested. Some signatures
could come from an expected ``neutron trapping'' effect.

\end{abstract}

\vskip -1.0cm
\section{Introduction}

Recently the development of new heavy ion facilities (radioactive beams) 
has driven 
the interest on the dynamical behaviour of asymmetric matter, see the
recent reviews \cite{bao,baranPR}. 
Here we focus our attention on relativistic heavy ion collisions, that
provide a unique terrestrial opportunity to probe the in-medium nuclear
interaction in high density and high momentum regions. 
An effective Lagrangian approach to the hadron interacting system is
extended to the isospin degree of freedom: within the same frame equilibrium
properties ($EoS$, \cite{qhd}) and transport dynamics 
\cite{KoPRL59,GiessenRPP56} can be consistently derived.

Within a covariant picture of the nuclear mean field, 
 for the description of the symmetry energy at saturation
($a_{4}$ parameter of the Weizs\"{a}ecker mass formula)
(a) only the Lorentz vector $\rho$ mesonic field, 
and (b) both, the vector $\rho$ (repulsive) and  scalar 
$\delta$ (attractive) effective 
fields \cite{liu,gait04} can be included. 
In the latter case the competition between scalar and vector fields leads
to a stiffer symmetry term at high density \cite{liu,baranPR}. We present
here observable effects, in fact enhanced, in the dynamics of heavy ion 
collisions. 
Here we focus our attention on collective isospin flows, in 
particular the elliptic ones, and on the isospin content of particle 
production,
in particular kaons. 
We finally show that in the compression stage of isospin asymmetric collisions
we can enter a mixed deconfined phase, if the $EoS$ conditions for the
existence of quark stars are met.

\section{Relativistic Transport}
The starting point is
a simple phenomenological version of the Non-Linear (with respect to the 
iso-scalar, Lorentz scalar $\sigma$ field) Walecka effective theory 
which corresponds 
to the 
Hartree or Relativistic Mean Field ($RMF$) approximation within the 
Quantum-Hadro-Dynamics \cite{qhd}. 
According to this model 
the presence of the hadronic medium leads to effective masses and 
momenta $M^{*}=M+\Sigma_{s}$,   
 $k^{*\mu}=k^{\mu}-\Sigma^{\mu}$, with
$\Sigma_{s},~\Sigma^{\mu}$
 scalar and vector self-energies. 
For asymmetric matter the self-energies are different for protons and 
neutrons, depending on the isovector meson contributions. 
We will call the 
corresponding models as $NL\rho$ and $NL\rho\delta$, respectively, and
just $NL$ the case without isovector interactions. 
For the more general $NL\rho\delta$ case  
the self-energies 
of protons and neutrons read:
\begin{equation}
\Sigma_{s}(p,n) = - f_{\sigma}\sigma(\rho_{s}) \pm f_{\delta}\rho_{s3}, 
~~~
\Sigma^{\mu}(p,n) = f_{\omega}j^{\mu} \mp f_{\rho}j^{\mu}_{3},~~
(upper~signs~for~neutrons),
\label{selfen}
\end{equation}
where $\rho_{s}=\rho_{sp}+\rho_{sn},~
j^{\alpha}=j^{\alpha}_{p}+j^{\alpha}_{n},\rho_{s3}=\rho_{sp}-\rho_{sn},
~j^{\alpha}_{3}=j^{\alpha}_{p}-j^{\alpha}_{n}$ are the total and 
isospin scalar 
densities and currents and $f_{\sigma,\omega,\rho,\delta}$  are the coupling 
constants of the various 
mesonic fields. 
$\sigma(\rho_{s})$ is the solution of the non linear 
equation for the $\sigma$ field \cite{liu,baranPR}.

For the description of heavy ion collisions we solve
the covariant transport equation of the Boltzmann type 
 \cite{KoPRL59,GiessenRPP56}  within the 
Relativistic Landau
Vlasov ($RLV$) method, using phase-space Gaussian test particles 
\cite{FuchsNPA589},
and applying
a Monte-Carlo procedure for the hard hadron collisions.
The collision term includes elastic and inelastic processes involving
the production/absorption of the $\Delta(1232 MeV)$ and $N^{*}(1440
MeV)$ resonances as well as their decays into pion channels,
 \cite{FeriniNPA762}.

It is worth to note that the nucleon mean field (Vlasov) propagation
is given by the following equations of motion for the test particles
trajectories \cite{baranPR}:
\begin{equation}
\frac{d}{d\tau}x_i^\mu=\frac{p^*_i(\tau)}{M^*_i(x)}~,~~~~~
\frac{d}{d\tau}p^{*\mu}_i=\frac{p^*_{i\nu}(\tau)}{M^*_i(x)}
{F}_i^{\mu\nu}\left(x_i(\tau)\right)+\partial^\mu M^*_i(x)~.
\label{eqmot}
\end{equation}
In order to have an idea of the dynamical effects of the
covariant nature of the interacting fields, we
write down, with some approximations, the ``force'' acting on a particle. 
Since we are interested in isospin contributions we will take into account 
only the isovector part of the interaction \cite{GrecoPLB562}:
\begin{equation}\label{force}
\frac{{d\vec p}^{\,*}_i}{d\tau}=\pm f_{\rho} \frac{p_{i\nu}}{M^*_i}
\left[\vec\nabla J_{3}^{\nu}-\partial^\nu\vec{J_3} \right]
\mp f_\delta \nabla \rho_{S3}
\approx 
\pm f_{\rho} \frac{E^*_i}{M^*_i}
\vec\nabla \rho_{3}
\mp f_\delta \vec\nabla \rho_{S3}, ~~(upper~signs~proton) 
\end{equation}
The Lorentz force (first term of Eq.(\ref{force})) shows a 
$\gamma=\frac{E^*_i}{M^*_i}$ boosting of the vector coupling, while 
from the second term we expect a $\gamma$-quenched $\delta$ contribution. 

\vskip -1.0cm
\section{Collective Flows}
The flow observables can be seen respectively as the
first and second coefficients of a Fourier expansion of the
azimuthal distribution \cite{OlliPRD46}:
$\frac{dN}{d\phi}(y,p_t) \approx 1+2V_1cos(\phi)+2V_2cos(2\phi)$
where $p_t=\sqrt{p_x^2+p_y^2}$ is the transverse momentum and $y$
the rapidity along beam direction. 
The transverse flow can be
expressed as: $V_1(y,p_t)=\langle \frac{p_x}{p_t} \rangle$.
The sideward (transverse) flow is a deflection of forwards and backwards 
moving particles, within the reaction plane. 
The second  coefficient of the expansion defines the elliptic
flow given by
$V_2(y,p_t)=\langle \frac{p_x^2-p_y^2}{p_t^2} \rangle$.
\begin{figure}
\begin{center}
 \includegraphics[angle=-90,scale=0.35]{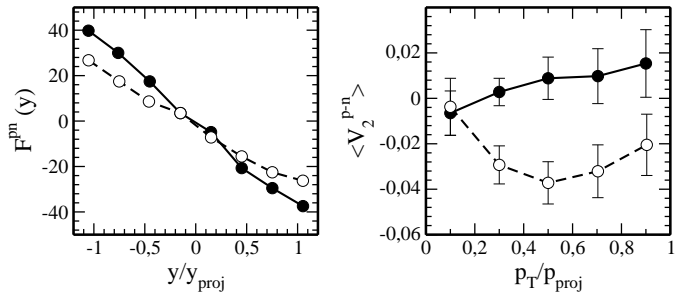}
\vskip -1.0cm
\caption{\small{Differential neutron-proton flows for the $^{132}Sn+^{124}Sn$
reaction at $1.5~AGeV$ ($b=6fm$) from the two different models for the
isovector mean fields.
Left: Transverse Flows. Right: Elliptic Flows.
Full circles and solid line: $NL\rho\delta$.
Open circles and dashed line: $NL\rho$.}
} 
\vskip -1.3cm
\label{flows}
\end{center}
\end{figure}
It measures the competition between in-plane and out-of-plane emissions. 
 The sign of $V_2$ indicates the azimuthal anisotropy of emission:
particles can be preferentially emitted either in the reaction
plane ($V_2>0$) or out-of-plane ($squeeze-out,~V_2<0$)
\cite{OlliPRD46,DanielNPA673}. 
For the isospin effects the neutron-proton $differential$ flows
$
V^{(n-p)}_{1,2} (y,p_t) \equiv V^n_{1,2}(y,p_t) - V^p_{1,2}(y,p_t)
$ 
have been suggested as very useful probes of the isovector part of 
the $EoS$ since they appear rather insensitive to the isoscalar potential
and to the in medium nuclear cross sections, \cite{BaoPRL82}.

In heavy-ion collisions around $1AGeV$ with
radioactive beams,
differential flows will directly exploit the
Lorentz nature of a scalar and a vector field, see the different
$\gamma$-boosting in the local force, Eq.(\ref{force}).
In Fig.\ref{flows}
transverse and elliptic differential flows are shown
for 
the $^{132}Sn+^{124}Sn$
reaction at $1.5~AGeV$ ($b=6fm$), 
 \cite{GrecoPLB562}. 
The effect of the different structure of the 
isovector channel is clear. Particularly evident is the splitting in 
the high $p_t$
region of the elliptic flow.
 In the $(\rho+\delta)$ dynamics the high-$p_t$ neutrons show a much larger 
$squeeze-out$.
This is fully consistent with an early emission (more spectator shadowing)
due to the larger $\rho$-field in the compression stage.
We expect similar effects, even enhanced, from 
the measurements of 
differential flows for light isobars, like $^3H~vs.~^3He$.

\vskip -1.0cm
\section{Isospin effects on sub-threshold kaon production at intermediate 
energies} 
Kaon production has been proven to be a reliable observable for the
high density $EoS$ in the isoscalar sector 
\cite{AichkoPRL55,FuchsPPNP56,HartPRL96}
Here we show that the $K^{0,+}$
production (in particular the $K^0/K^+$ yield ratio) can be also used to 
probe the isovector part of the $EoS$.

Using our $RMF$ transport approach  we analyze 
pion and kaon production in central $^{197}Au+^{197}Au$ collisions in 
the $0.8-1.8~AGeV$
 beam 
energy range, comparing models giving the same ``soft'' $EoS$ for symmetric 
matter and with different effective field choices for 
$E_{sym}$. We will use three Lagrangians with constant 
nucleon-meson 
couplings ($NL...$ type, see before) and one with density
dependent couplings ($DDF$, see \cite{gait04}), recently suggested 
for better nucleonic properties of neutron stars \cite{Klahn06}.
In the $DDF$ model
the $f_{\rho}$ is exponentially decreasing with density, resulting in a 
rather "soft" 
symmetry term at high density. 
The hadron mean field propagation, which goes beyond the 
``collision cascade'' picture, is
essential for particle production yields: in particular the
isospin dependence of the self-energies directly affects the
energy balance of the inelastic channels.

\begin{figure}[t] 
\begin{center}
\includegraphics[scale=0.28]{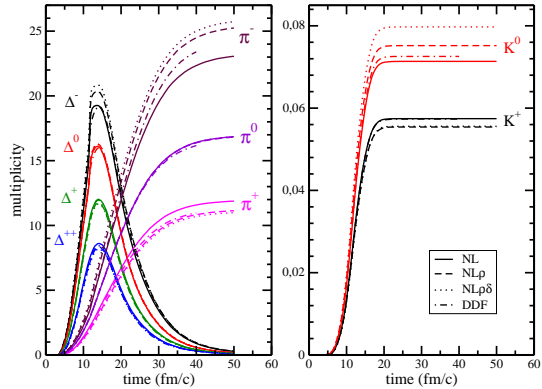} 
\vskip -1.0cm
\caption{\small{Time evolution of the $\Delta^{\pm,0,++}$ resonances
and pions $\pi^{\pm,0}$ 
(left),  and  kaons ($K^{+,0}$
 (right) for a central ($b=0$ fm impact parameter)  
Au+Au collision at 1 AGeV incident energy. Transport calculation using the  
$NL, NL\rho, NL\rho\delta$ and $DDF$ models for the iso-vector part of the  
nuclear $EoS$ are shown.}  
}
\vskip -1.3cm
\label{kaon1} 
\end{center}
\end{figure}

Fig. \ref{kaon1} reports  the temporal evolution of $\Delta^{\pm,0,++}$  
resonances, pions ($\pi^{\pm,0}$) and kaons ($K^{+,0}$)  
for central Au+Au collisions at $1AGeV$.
It is clear that, while the pion yield freezes out at times of the order of 
$50 fm/c$, i.e. at the final stage of the reaction (and at low densities),
kaon production occur within the very early (compression) stage,
 and the yield saturates at around $20 fm/c$. 
From Fig. \ref{kaon1} we see that the pion results are  
weakly dependent on the  
isospin part of the nuclear mean field.
However, a slight increase (decrease) in the $\pi^{-}$ ($\pi^{+}$) 
multiplicity is observed when going from the $NL$ (or $DDF$) to the 
$NL\rho$ and then to
the $NL\rho\delta$ model, i.e. increasing the vector contribution $f_\rho$
in the isovector channel. This trend is 
more pronounced for kaons, see the
right panel, due to the high density selection of the source and the
proximity to the production threshold. 

When isovector fields are included the symmetry potential energy in 
neutron-rich matter is repulsive for neutrons and attractive for protons.
In a $HIC$ this leads to a fast, pre-equilibrium, emission of neutrons.
 Such a $mean~field$ mechanism, often referred to as isospin fractionation
\cite{bao,baranPR}, is responsible for a reduction of the neutron
to proton ratio during the high density phase, with direct consequences
on particle production in inelastic $NN$ collisions.

$Threshold$ effects represent a more subtle point. The energy 
conservation in
a hadron collision in general has to be formulated in terms of the canonical
momenta, i.e. for a reaction $1+2 \rightarrow 3+4$ as
$
s_{in} = (k_1^\mu + k_2^\mu)^2 = (k_3^\mu + k_4^\mu)^2 = s_{out}.
$
Since hadrons are propagating with effective (kinetic) momenta and masses,
 an equivalent relation should be formulated starting from the effective
in-medium quantities $k^{*\mu}=k^\mu-\Sigma^\mu$ and $m^*=m+\Sigma_s$, where
$\Sigma_s$ and $\Sigma^\mu$ are the scalar and vector self-energies,
Eqs.(\ref{selfen}).
The self-energy contributions will influence the particle production at the
level of thresholds as well as of the phase space available in the final 
channel

In neutron-rich colliding systems {\it Mean field} 
and {\it threshold} effects
are acting in opposite directions on particle production  and might 
compensate each other.
 As an example, $nn$
collisions excite $\Delta^{-,0}$ resonances which decay mainly to $\pi^-$.
 In a
neutron-rich matter the mean field effect pushes out neutrons making the 
matter more symmetric and thus decreasing the $\pi^-$ yield. The threshold 
effect on the other hand is increasing the rate of $\pi^-$'s due to the
enhanced production of the $\Delta^-$ resonances: 
now the $nn \rightarrow p\Delta^-$ process is favored
(with respect to $pp \rightarrow n\Delta^{++}$) 
 since more effectively a neutron is converted into a proton.
Such interplay between the two mechanisms cannot be 
fully included in a non-relativistic dynamics,
in particular in calculations where the baryon symmetry potential is
treated classically \cite{BaoPRC71,QLiPRC72}.

We have to note that in a previous study of kaon production in excited nuclear
matter the dependence of the $K^{0}/K^{+}$ yield ratio on the effective
isovector interaction appears much larger (see Fig.8 of 
ref.\cite{FeriniNPA762}).
The point is that in the non-equilibrium case of a heavy ion collision
the asymmetry of the source where kaons are produced is in fact reduced
by the $n \rightarrow p$ ``transformation'', due to the favored 
$nn \rightarrow p\Delta^-$ processes. This effect is almost absent at 
equilibrium due to the inverse transitions, see Fig.3 of 
ref.\cite{FeriniNPA762}. Moreover in infinite nuclear matter even the fast
neutron emission is not present. 
This result clearly shows that chemical equilibrium models can lead to
uncorrect results when used for transient states of an $open$ system.

\section{Testing Deconfinement at High Isospin Density}
The hadronic matter is expected to undergo a phase transition 
into a deconfined phase of quarks and gluons at large densities 
and/or high temperatures. On very general grounds,
the transition's critical densities are expected to depend
on the isospin of the system, but no experimental tests of this 
dependence have been performed so far.
Moreover, up to now, data on the phase transition have been 
extracted from
ultrarelativistic collisions, when large temperatures but low baryon densities
are reached. 
\begin{figure}
\begin{center}
\includegraphics[scale=0.36]{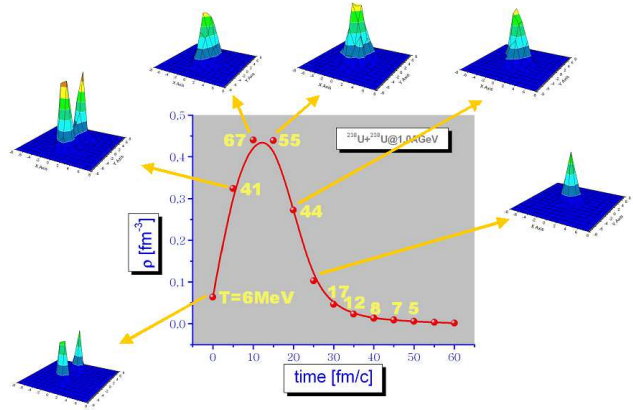}
\vskip -1.3cm
\caption{
\small{$^{238}U+^{238}U$, $1~AGeV$, semicentral. Correlation between 
density, 
temperature, momentum
thermalization inside a cubic cell, 2.5 $fm$ wide, in the center
of mass of the system.}  
}
\label{figUU}
\vskip -1.3cm
\end{center}
\end{figure}
In order to check the possibility of observing some precursor signals
of some new physics even in collisions of stable nuclei at
intermediate energies we have performed some event simulations for the
collision of very heavy, neutron-rich, elements. We have chosen the
reaction $^{238}U+^{238}U$ (average proton fraction $Z/A=0.39$) at
$1~AGeV$ and semicentral impact parameter $b=7~fm$ just to increase
the neutron excess in the interacting region. 

In  Fig.~\ref{figUU} we report the evolution of momentum distribution
and baryon density in a space cell located in the c.m. of the system.
We see that after about $10~fm/c$ a nice local
equilibration is achieved.  We have a unique Fermi distribution and
from a simple fit we can evaluate the local temperature.
We note that a rather exotic nuclear matter is formed in a transient
time of the order of $10~fm/c$, with baryon density around $3-4\rho_0$,
temperature $50-60~MeV$, energy density $500~MeV~fm^{-3}$ and proton
fraction between $0.35$ and $0.40$, likely inside the estimated mixed 
phase region, see the following..

Here we study the isospin dependence of the transition densities
 \cite{MuellerNPA618} in a systematic way,
 exploring also
the possibility of forming a mixed-phase of quarks and hadrons in
experiments at energies of the order of a few $GeV$ per nucleon.
Concerning the hadronic phase, we use the relativistic
non-linear model of Glendenning-Moszkowski (in particular the ``soft''
$GM3$ choice) 
\cite{GlendenningPRL18}, where the isovector part is treated 
just with $\rho$ meson coupling, and
the iso-stiffer $NL\rho\delta$ interaction \cite{deconf06}. 
For the quark phase we consider the $MIT$ bag model \cite{MitbagPRD9}
with various bag pressure constants.  In particular 
we are interested in those parameter sets
which would allow the existence of quark stars
\cite{HaenselAA160,DragoPLB511}, i.e. parameters sets for
which the so-called Witten-Bodmer hypothesis is satisfied
\cite{WittenPRD30,BodmerPRD4}. 
One of the
aim of our work it to show that if quark stars are indeed possible,
it is then very likely to find signals of the formation of a mixed
quark-hadron phase in intermediate-energy heavy-ion experiments
\cite{deconf06}.

The structure of the mixed phase is obtained by
imposing the Gibbs conditions \cite{Landaustat,GlendenningPRD46} for
chemical potentials and pressure and by requiring
the conservation of the total baryon and isospin densities
\begin{eqnarray}\label{gibbs}
&&\mu_B^{(H)} = \mu_B^{(Q)}\, ,~~  
\mu_3^{(H)} = \mu_3^{(Q)} \, ,  
~~~P^{(H)}(T,\mu_{B,3}^{(H)}) = P^{(Q)} (T,\mu_{B,3}^{(Q)})\, ,\nonumber \\
&&\rho_B=(1-\chi)\rho_B^H+\chi\rho_B^Q \, ,
~~\rho_3=(1-\chi)\rho_3^H+\chi\rho_3^Q\, , 
\end{eqnarray}
where $\chi$ is the fraction of quark matter in the mixed phase.
In this way we get the $binodal$ surface which gives the phase coexistence 
region
in the $(T,\rho_B,\rho_3)$ space
\cite{GlendenningPRD46,MuellerNPA618}. For a fixed value of the
conserved charge $\rho_3$ 
 we will study the boundaries of the mixed phase
region in the $(T,\rho_B)$ plane. 
In the hadronic phase the charge chemical potential is given by
$
\mu_3 = 2 E_{sym}(\rho_B) \frac{\rho_3}{\rho_B}\, .
$ 
Thus, we expect critical densities
rather sensitive to the isovector channel in the hadronic $EoS$.

\begin{figure}
\begin{center}
\includegraphics[angle=+90,scale=0.37]{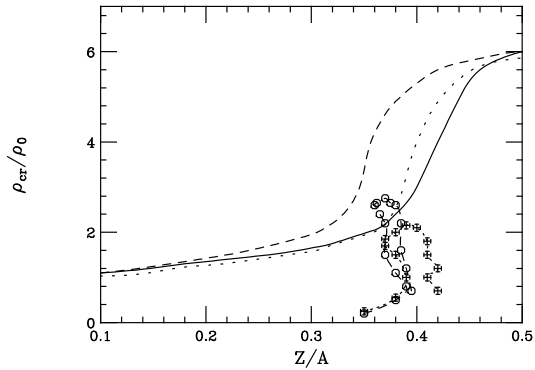}
\vskip -1.0cm
\caption{
\small{Variation of the transition density with proton fraction for various
hadronic $EoS$ parameterizations. Dotted line: $GM3$ parametrization;
 dashed line: $NL\rho$ parametrization; solid line: $NL\rho\delta$ 
parametrization. For the quark $EoS$, the $MIT$ bag model with
$B^{1/4}$=150 $MeV$.
The points represent the path followed
in the interaction zone during a semi-central $^{132}$Sn+$^{132}$Sn
collision at $1~AGeV$ (circles) and at $300~AMeV$ (crosses). 
}}
\vskip -1.3cm
\label{rhodelta}
\end{center}
\end{figure}

In Fig.~\ref{rhodelta}  we show the crossing
density $\rho_{cr}$ separating nuclear matter from the quark-nucleon
mixed phase, as a function of the proton fraction $Z/A$.  
We can see the effect of the
$\delta$-coupling towards an $earlier$ crossing due to the larger
symmetry repulsion at high baryon densities.
In the same figure we report the paths in the $(\rho,Z/A)$
plane followed in the c.m. region during the collision of the n-rich
 $^{132}$Sn+$^{132}$Sn system, at different energies. At
$300~AMeV$ we are just reaching the border of the mixed phase, and we are
well inside it at $1~AGeV$. 
Statistical
fluctuations could help in  reducing the density at which drops of quark
matter form. The reason is that a small bubble can
be energetically favored if it contains quarks whose Z/A ratio is
{\it smaller} than the average value of the surrounding region. This
is again due to the strong Z/A dependence of the free energy, which
favors clusters having a small electric charge. 
Moreover, since 
fluctuations favor the formation of bubbles having a smaller Z/A,
neutron emission from the central collision area should be suppressed,
which could give origin to specific signatures of the mechanism
described in this paper. This corresponds to a {\it neutron trapping}
effect, supported also by a symmetry energy difference in the
two phases.
In fact while in the hadron phase we have a large neutron
potential repulsion (in particular in the $NL\rho\delta$ case), in the
quark phase we only have the much smaller kinetic contribution.
 If in a
pure hadronic phase neutrons are quickly emitted or ``transformed'' in
protons by inelastic collisions, when the mixed phase
starts forming, neutrons are kept in the interacting system up to the
subsequent hadronization in the expansion stage \cite{deconf06}.
Observables related to such neutron ``trapping'' could be an
inversion in the trend of the formation of neutron rich fragments
and/or of the $\pi^-/\pi^+$, $K^0/K^+$ yield ratios for reaction
products coming from high density regions, i.e. with large transverse
momenta.  In general we would expect a modification of the rapidity
distribution of the emitted ``isospin'', with an enhancement at
mid-rapidity joint to large event by event fluctuations..

\section{Perspectives}
We have shown that collisions of n-rich heavy ions at intermediate energies
can bring new information on the isovector part of the in-medium interaction
at high baryon densities, qualitatively different from equilibrium
$EoS$ properties. We have presented quantitative results for differential
collective flows and yields of charged pion and kaon ratios.
Important non-equilibrium effects for particle production are stressed.
Finally our study supports the possibility of observing
precursor signals of the phase transition to a mixed hadron-quark matter
at high baryon density in the collision, central or semi-central, of
neutron-rich heavy ions in the energy range of a few $GeV$ per
nucleon.  As signatures we 
suggest to look at observables particularly sensitive to the
expected different isospin content of the two phases, which leads to a
neutron trapping in the quark clusters.
The isospin structure of hadrons produced at high transverse momentum
should be a good indicator of the effect. 

In conclusion the results presented here appear very promising for 
the possibility
of extracting information from terrestrial laboratories on the
Lorentz structure of the isovector nuclear interaction in a medium
at densities of astrophysical interest. The use of radioactive beams
at relativistic energies would be extremely important.
\vskip 0.5cm
{\it Acknowledgements}

\noindent 
We warmly thanks A.Drago and A.Lavagno for the intense 
collaboration on the
mixed hadron-quark phase transition at high baryon and isospin density.


\end{document}